\documentclass[twocolumn,pra,superscriptaddress,preprintnumbers,amsmath,amssymb]{revtex4-1}

\usepackage{ae,aecompl}

\usepackage{graphicx}
\usepackage[normalem]{ulem}
\usepackage{soul,xcolor}
\setstcolor{red}

\usepackage[latin1]{inputenc}
\usepackage[T1]{fontenc}

\usepackage{color}

\newcommand{\etal}{{\it et al.}}

\begin{document}

\def\pss#1#2#3{Phys.~Stat.~Sol.~{\bf #1},\ #2\ (#3)}
\def\apl#1#2#3{Appl.~Phys.~Lett.~{\bf #1},\ #2\ (#3)}
\def\jpb#1#2#3{J.~Phys.~B:~{\bf #1},\ #2\ (#3)}
\def\jpc#1#2#3{J.~Phys.~Chem.~{\bf #1},\ #2\ (#3)}
\def\jpcb#1#2#3{J.~Phys.~Chem. B~{\bf #1},\ #2\ (#3)}
\def\jcp#1#2#3{J.~Chem.~Phys.~{\bf #1},\ #2\ (#3)}
\def\cpl#1#2#3{Chem.~Phys.~Lett.~{\bf #1},\ #2\ (#3)}
\def\pr#1#2#3{Phys.~Rev~{\bf #1},\ #2\ (#3)}
\def\pra#1#2#3{Phys.~Rev.~A~{\bf #1},\ #2\ (#3)}
\def\prb#1#2#3{Phys.~Rev.~B~{\bf #1},\ #2\ (#3)}
\def\prc#1#2#3{Phys.~Rev.~C~{\bf #1},\ #2\ (#3)}
\def\prd#1#2#3{Phys.~Rev.~D~{\bf #1},\ #2\ (#3)}
\def\pre#1#2#3{Phys.~Rev.~E~{\bf #1},\ #2\ (#3)}
\def\rmp#1#2#3{Rev.~Mod.~Phys.~{\bf #1},\ #2\ (#3)}
\def\prl#1#2#3{Phys.~Rev.~Lett.~{\bf #1},\ #2\ (#3)}
\def\sci#1#2#3{Science~{\bf #1},\ #2\ (#3)}
\def\nat#1#2#3{Nature~{\bf #1},\ #2\ (#3)}
\def\apj#1#2#3{Astrophys.~J.~{\bf #1},\ #2\ (#3)}

\def\bea{\begin{eqnarray}}
\def\eea{\end{eqnarray}}
\def\be{\begin{equation}}
\def\ee{\end{equation}}
\def\etal{{\it et al.}}

\newcommand{\JCP}[1]{J. Chem. Phys. {\bf #1}}
\newcommand{\ApJ}[1]{Astrophys. J. {\bf#1}}
\newcommand{\CPL}[1]{Chem. Phys. Lett. {\bf #1}}
\newcommand{\PRA}[1]{Phys. Rev. A {\bf #1}}
\newcommand{\JPC}[1]{J. Phys. Chem. {\bf #1}}
\newcommand{\JPCA}[1]{J. Phys. Chem. A {\bf #1}}
\newcommand{\PRL}[1]{Phys. Rev. Lett. {\bf #1}}
\def\BE{\begin{equation}} \def\EE{\end{equation}}

\preprint{Preprint}

\title{Electric-field noise from carbon-adatom diffusion on a Au(110) surface:
 first-principles calculations and experiments}
\date{\today}
\author{E. Kim}
\affiliation{Department of Physics and Astronomy, University of Nevada, Las Vegas, NV 89154-4002}
\author{A. Safavi-Naini}
\affiliation{JILA, 440 University Avenue, Boulder, Colorado  80302}
\author{D. A. Hite}
\affiliation{NIST, 325 Broadway, Boulder, Colorado 80305}
\author{K. S. McKay}
\affiliation{NIST, 325 Broadway, Boulder, Colorado 80305}
\author{D. P. Pappas}
\affiliation{NIST, 325 Broadway, Boulder, Colorado 80305}
\author{P. F. Weck}
\affiliation{Sandia National Laboratories, P.O. Box 5800, Albuquerque, New Mexico 87185-0779}
\author{H. R. Sadeghpour}
\affiliation{ITAMP, Harvard-Smithsonian Center for Astrophysics, Cambridge, Massachusetts 02138}
%

\begin{abstract}
The decoherence of trapped-ion quantum gates due to heating of their
 motional modes is a fundamental science and engineering
 problem. This heating is attributed to electric-field noise arising from the trap-electrode surfaces.
 In this work, we investigate the source of this noise
 by focusing on the diffusion of carbon-containing adsorbates on the
 surface of Au(110). We show by density
 functional theory, based on detailed scanning probe microscopy, how the carbon adatom diffusion on the gold surface
 changes the energy landscape, and how the adatom dipole moment varies
 with the diffusive motion. A simple model for the diffusion noise,
 which varies quadratically with the variation of the dipole moment,
 qualitatively reproduces the measured noise spectrum, and the estimate
 of the noise spectral density is in accord with measured values.
\end{abstract}
\date{\today}
\maketitle

\section{Introduction}


Trapped ions are a promising platform for demonstrating coherent operations for quantum information applications \cite {Monroe2013}, however heating of their motional modes remains a major obstacle to continued progress \cite {Hite2013}. In particular, motional heating caused by electric-field noise originating from the trap-electrode surfaces has proven to be a difficult problem to mitigate, ever since it was first observed more than two decades ago.  This decoherence mechanism scales strongly with the distance of the ion to the nearest electrode, and therefore is a barrier to scalability through miniaturization.

The origin of this noise source has been suspected to arise from surface processes, based on experimental evidence on scaling with ion-electrode distance \cite{Turchette00, Deslauriers06}, electrode temperature \cite{Labaziewicz08b, Chiaverini14}, and spectral-density frequency dependence \cite{Turchette00, Deslauriers04, Deslauriers06, Labaziewicz08b, Epstein07, Allcock11}.  The surface origin of the noise was recently confirmed experimentally upon in situ surface treatment by ion bombardment, demonstrating a reduction in motional heating by more than two orders of magnitude \cite{Hite12, Daniilidis14}.  It has been suggested that this reduction in heating is related to the removal of surface contamination. One proposal has modeled the noise with thermally activated, normal-to-the-surface fluctuating adsorbed dipoles, and obtained a noise spectrum of the same magnitude as that observed in experiments \cite{Safavi-Naini11}.  The exact mechanism that gives rise to electric-field noise at the location of the ion, however, still remains elusive.

In this joint theoretical and experimental work, we investigate electric-field noise due to diffusion of carbon adatoms on gold surfaces.  We are motivated by the observation that carbon is a dominant contaminant on gold trap-electrode surfaces \cite{Hite12}, and that ordered Au(110)-like structures are observed in scanning tunneling microscopy (STM) measurements after trap electrodes have been treated with ion bombardment(cf. Sec. IV).  Density functional theory (DFT) simulations provide the first detailed values for the energy and dipole landscape of the adsorbed carbon atoms on a Au(110) surface, and are used subsequently in an analytical derivation of noise due to classical diffusion.  The model shows how the electric-field noise varies in a non-monotonic fashion as a function of the degree of carbon-adatom coverage.  The electric-field noise spectral density is a function of the variation in the adatom dipole moment, the surface diffusion constant, and patch size with different work functions.  Using realistic parameters, the theory predicts an electric-field noise spectral density consistent with experimental measurement.

\section{Experimental Method} \label{motivation}
The DFT calculations of submonolayer coverages of the C/Au(110) system are motivated by the experimental findings that have indicated a non-monotonic behavior of the electric-field noise within this coverage regime.  In the experiments, we employ a stylus-type Paul trap with room temperature electrodes, similar to the trap in Ref. \cite{Arrington13}.  We confine a $^{25}$Mg$^+$ ion 63 $\mu$m above the nearest electrode, and measure heating rates of a 4.7-MHz motional mode as a function of cumulative doses of ion bombardment. The repeated treatments incrementally remove the contaminants providing access to submonolayer coverages.

The electric-field noise spectral density $S_E (\omega)$ and the heating rate in terms of rate of increase in motional quanta, $\dot{\overline{n}}$ $\equiv$ \textit{d$\overline{n}$/dt}, are related by:

\begin{equation}
S_E(\omega) = {4m \hbar \omega \over q^2} \dot{\overline{n}},
\label{eq:spectral noise}
\end{equation}
where $\omega/2\pi$ is the motional frequency of the ion in the trap, \textit{m} is the ion mass, \textit{q} is its charge, and $\hbar$ is the reduced Planck's constant \cite{Turchette00}.

We begin the set of measurements with a microfabricated trap chip consisting of electroplated Au.   As fabricated, these surfaces are typically covered with 2 - 3 monolayers (ML) of carbon contamination, as measured by Auger electron spectroscopy (AES) \cite{Footnote1}. The adventitious adsorption on the gold electrode surfaces likely
originates from atmospheric hydrocarbons. The carbon AES line shapes are
consistent with graphitic-like adsorbates, and often do not show oxygen
peaks \cite{Hite12, 2014arxiv}. In this theoretical work, and for computational economy,  we consider
only carbon adsorbates.  Future
theoretical and experimental efforts will investigate the presence of
more complex molecular hydrocarbon adsorbates.

To achieve various submonolayer coverages of the contamination, we incrementally dose the trap electrode surfaces with Ar$^+$-bombardment, where we define the total energy dose in J/cm$^2$ by

\begin{equation}
E = {V j t},
\label{eq:total energy dose}
\end{equation}
where $V$ is the ion-beam acceleration voltage in volts, $j$ is the ion-beam current density in A/cm$^2$, and $t$ is the duration of the treatment in seconds.  The current density of the ion beam was calibrated using a Faraday cup with a 0.5 mm aperture in a separate system.  Based on many depth-profiling measurements using AES in a separate system with duplicate trap chips, we can infer approximate coverages that result from the various treatments to the ion trap.

In Fig.1, the electric-field noise spectral density is seen to increase by an order of magnitude, and peaks at approximately $1/2$ ML of carbon.  At an estimated coverage of $\sim$ 0.1 ML, the noise level drops by roughly two orders of magnitude, followed by another treatment that removed the carbon to undetectable levels ($<$ 0.05 ML) as determined by AES measurements.   These data indicate that in the submonolayer coverage regime, electric-field noise from surfaces behaves in a non-monotonic manner, and that trapped ions are sensitive to very low concentrations of adsorbates.  This non-monotonic behavior is reminiscent of surface-diffusion experiments in which the surface-diffusion parameter $D$ was observed to vary by orders of magnitude as a function of coverage in the submonolayer regime, with a characteristic peak at $\sim$ 0.5 ML \cite {Butz77, Asada80, Snabl98, Naumovets05}.

\begin{figure}[h!]
\begin{center}
\includegraphics[width=0.45\textwidth]{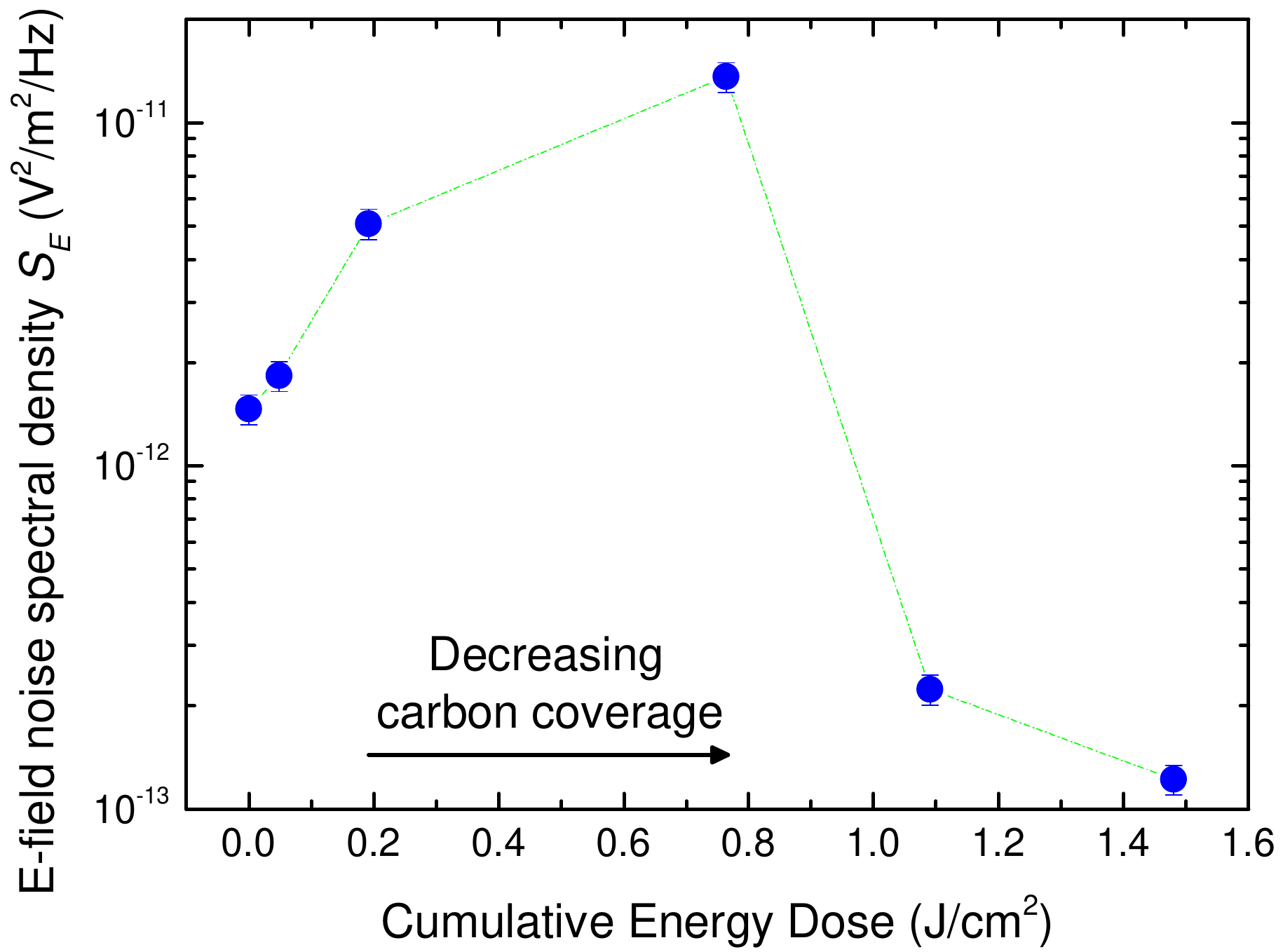}
\end{center}
\caption{\label{fig1-noisedata} Electric-field noise spectral density at $\omega/2\pi$ = 4.7 MHz vs. the cumulative dose for incremental treatments.  $S_E$ is determined from Eq. 1 and the motional heating rate of a $^{25}$Mg$^+$ ion trapped 63 $\mu$m above the nearest electrode.  The ion-bombardment treatments to the electrodes employ 500 eV argon ions with typical current densities of $\sim$ 0.2 - 0.5 $\mu$A/cm$^2$.  As the treatments proceed, the coverage of the contamination decreases.  The line connecting the points is intended only to indicate sequential data, not to indicate a trend in the behavior. This non-monotonic behavior in the noise with successive treatments has been observed in each of the six ion traps we have investigated with ion-bombardment treatments.
}
\end{figure}

\section{Computational Method} \label{methods}

The theoretical study begins with a characterization of the C-adatom motion on a Au(110) surface.
Total-energy calculations of bulk Au and Au(110) surfaces, with and without
carbon adsorbate atoms, were performed using spin-polarized density
functional theory (DFT) as implemented in the Vienna Ab initio Software Package
(VASP) \cite{kresse1996}. The exchange-correlation energy was calculated
using the generalized gradient approximation (GGA) \cite{perdew1992gga} with
the parameterization of Perdew, Burke, and Ernzerhof (PBE) \cite{perdew1996}.
The interaction between valence electrons and ionic cores was described
by the Projector Augmented Wave (PAW) method \cite{blochl1994,kresse1999}.
The Au $5d^{10}6s^1$ and C $2s^22p^2$ electrons were treated
explicitly as valence electrons in the Kohn-Sham (KS) equations and the
remaining cores were represented by PAW pseudopotentials. The KS
equations were solved using the blocked Davidson iterative matrix
diagonalization scheme followed by the residual vector minimization
method. The plane-wave cutoff energy for the electronic wavefunctions
was set to 500~eV.

All structures were optimized with periodic boundary conditions
applied using the conjugate gradient method, accelerated using
the Methfessel-Paxton Fermi-level smearing \cite{methfessel1989}
with a width of 0.2~eV. The total energy of the system and
Hellmann-Feynman forces acting on atoms were calculated with
convergence tolerances set to $10^{-3}$~eV and 0.01~eV/{\AA},
respectively. Structural optimizations and properties calculations
were carried out using the Monkhorst-Pack special $k$-point
scheme \cite{monkhorst1976} with $11\times11\times11$ and
$5\times5\times1$ meshes for integrations in the Brillouin
zone (BZ) of bulk and slab systems, respectively.

A $(2 \times 2)-$periodic supercell slab was constructed by cleaving
relaxed bulk Au with lattice constant 4.14 \AA, i.e. in close agreement
with the experimental value of 4.0780~\AA~at $25^{\circ}$C \cite{dutta1963}.
The slab model consisted of six-layer thick Au$(110)$ with the
reconstructed $(2 \times 1)$ superstructure.  The $(2 \times 1)$ reconstruction on Au(110) is called the ``missing row" structure
because every second row of the $\left<1\overline{1}0\right>$ surface chains is missing,
as observed in STM experiments
(cf. Fig. \ref{fig2-STM}).
The top four layers, on the side of the slab used to model atom adsorprtion,
were allowed to relax while the bottom two layers were kept fixed to mimic
the bulk structure.
Although a large vacuum region ($\simeq 15$~\AA~) was used between
periodic slabs, the creation of dipoles upon adsorption of atoms on
only one side of the slab can lead to spurious interactions
between the dipoles of successive slabs. In order to circumvent
this problem, a dipole correction was applied by means of a dipole
layer placed in the vacuum region following the method outlined
by Neugebauer and Scheffler \cite{neugebauer1992}.

\section{Results and Discussion}

\subsection{Au$(110)$-$(2\times 1)$ superstructure}

\begin{figure}[b]
\begin{center}
\includegraphics[width=0.5\textwidth]{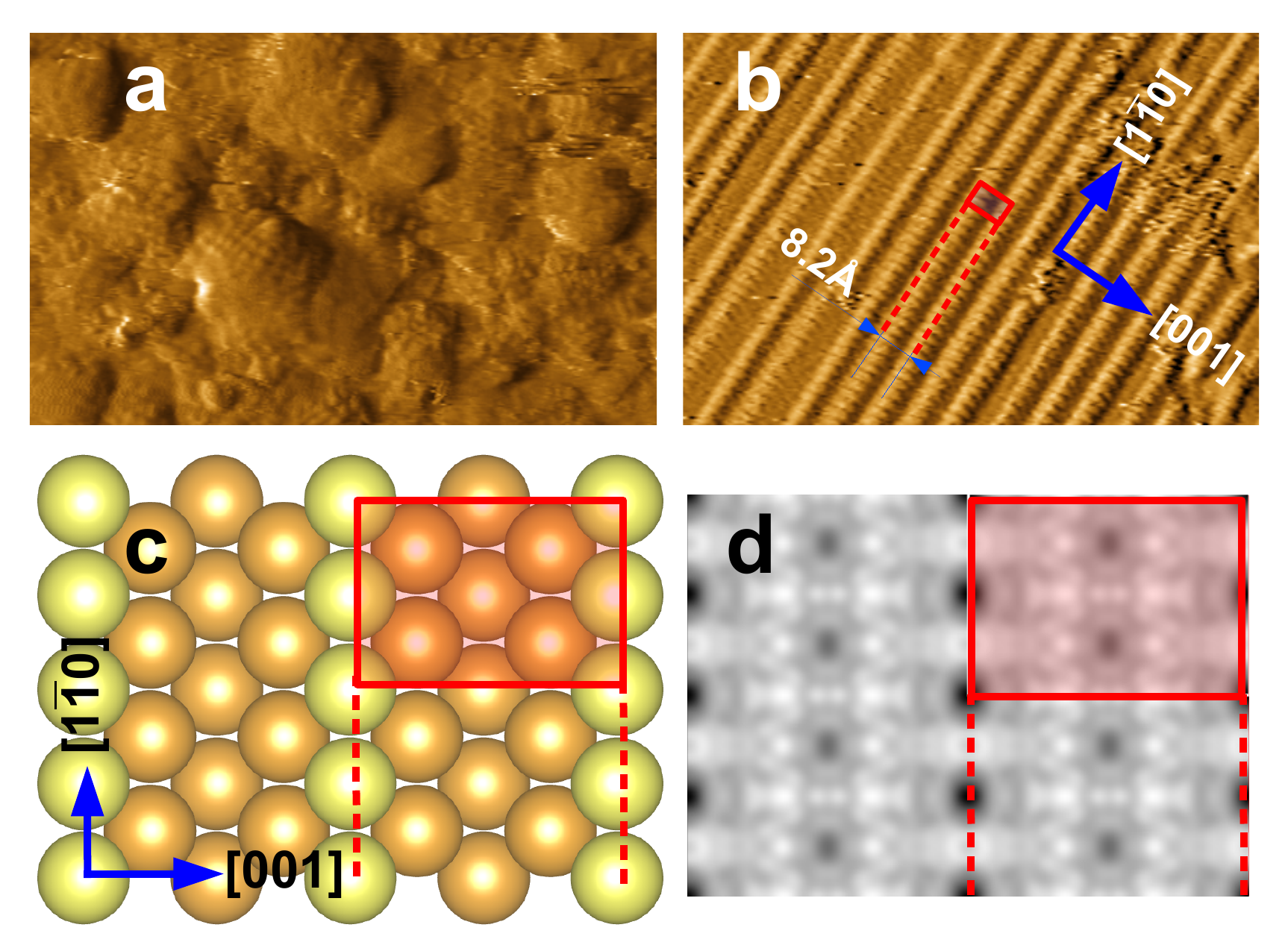}
\end{center}
\caption{\label{fig2-STM} (a) Derivative STM image (17 nm $\times$ 10 nm) of an untreated ion-trap surface covered by $\sim$ 2.5 ML of adventitious carbon.  (b) Derivative STM image (17 nm $\times$ 10 nm) of same trap after treatment by Ar$^+$ bombardment revealing a clean Au$(110)$ surface. (c) Top view of the Au$(110)$-$(2\times 1)$
reconstructed superstructure relaxed with DFT (top-row Au surface atoms
are shown in light yellow) and (d) its simulated STM image.
The $(2\times 2)$ simulation supercell along the $[1\overline{1}0]$ and
$[001]$ directions is represented by a red rectangle, and the
$\left<1\overline{1}0\right>$ rows are indicated by red dashed lines.  The reconstructed $(2\times 1)$ structure
features alternately missing $\left<1\overline{1}0\right>$ rows.}
\end{figure}

The trap electrodes used in our experiments are microfabricated with electroplated Au.  Before treatment by in situ ion bombardment, the surfaces are typically covered by 2 - 3 ML of carbon contamination as measured by AES.  Figure \ref{fig2-STM} (a) shows an STM image of such a contaminated surface, characterized by a clustered and disordered morphology, where the full height scale is $\sim$ 2 nm in topography.  This surface corresponds to a trap producing high electric-field noise, as typically measured in ion-trap heating-rate measurements from untreated electrode surfaces (see \cite {Hite2013} and references therein).

After an ion-bombardment treatment sufficient to remove the carbon contamination as confirmed by AES, STM measurements reveal ordered Au(110)-(2 $\times$ 1)-like structures on the trap chip (Fig. \ref{fig2-STM} (b)).  This would correspond to a trap with low electric-field noise \cite {Hite12, Daniilidis14}.  These treatments do not include post annealing, and therefore the treated surfaces have a rough, hill-and-valley morphology on the tens of nanometer scale.  The full height for the image in Fig. \ref{fig2-STM} (b) is $\sim$ 1 nm in topography.  In other trap-electrode samples treated with ion bombardment, we have also observed Au(100)-like and Au(111)-like structures consistent with the polycrystalline structure of the electroplated Au electrodes.  We focus here on the Au(110)-(2 $\times$ 1) surface and compute the dipole variation of carbon adsorbates diffusing across the surface.

The reconstructed $(2 \times 1)$ superstructure features alternately
missing $\left<1\overline{1}0\right>$ rows [Fig. \ref{fig2-STM}(b)], as
contrasted by the high-temperature $(1 \times 1)$ bulk-like structure.
Previous studies have indicated the occurrence of a continuous surface
order-disorder transition of the 2-D Ising universality class between
the $(2 \times 1)$ and $(1 \times 1)$ structures on
Au$(110)$ \cite{nguyen1988}.
For the sake of comparison, a STM image was simulated in a plane
$\simeq 1$~\AA~above the surface atoms of the Au$(110)$-$(2\times 1)$
slab model optimized with DFT [Fig. \ref{fig2-STM}(c)].
Good agreement is found between observed and simulated STM
images [Figs. \ref{fig2-STM}(b) and (d)], with adjacent
$\left<1\overline{1}0\right>$ top rows separated by
$\simeq 8.2$~\AA~along the $[001]$ direction.

\subsection{Carbon adsorption on Au$(110)$}

\begin{figure}[t]
\begin{center}
\includegraphics[width=0.5\textwidth]{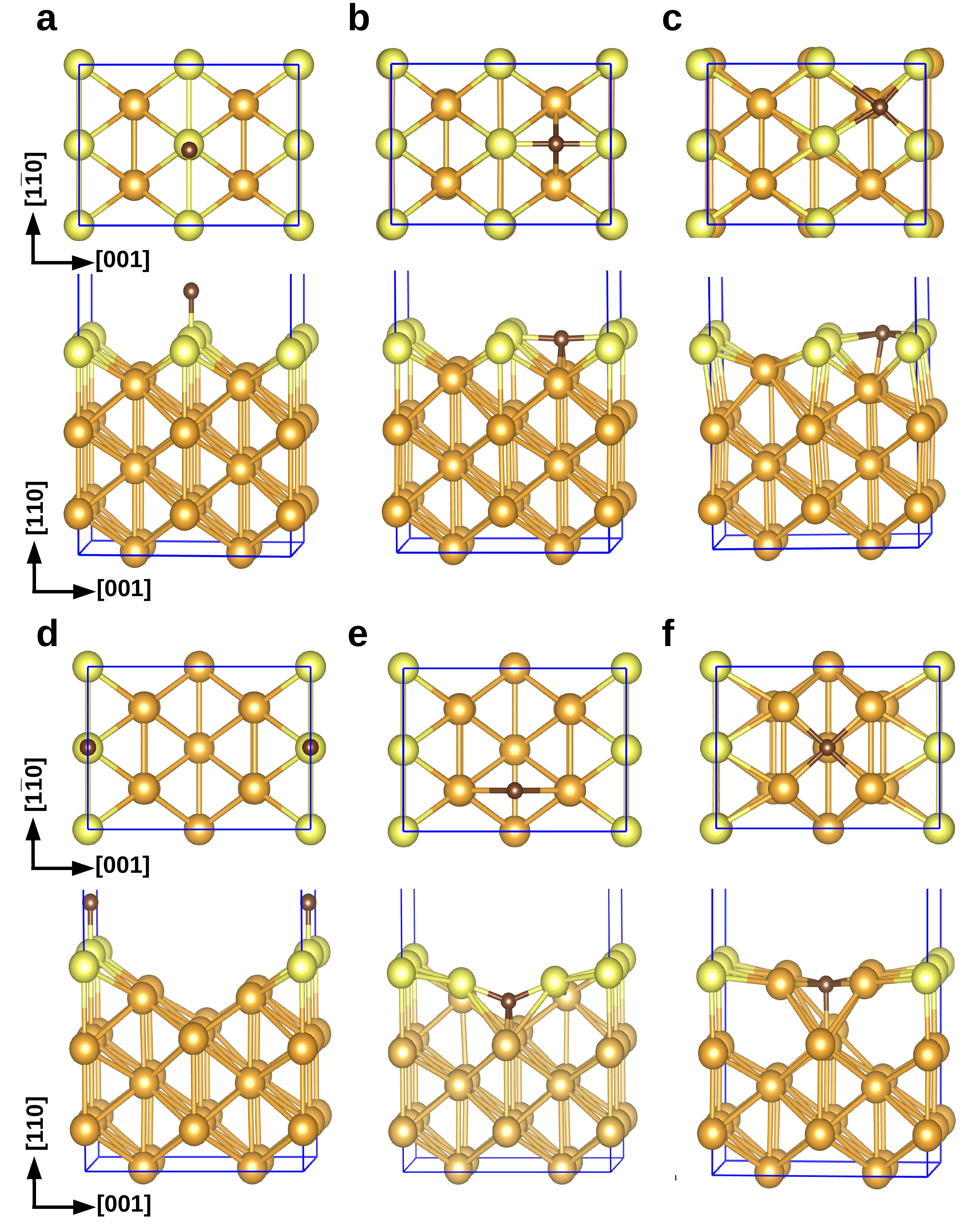}
\end{center}
\caption{\label{fig3}
Carbon atom adsorption ($\theta=0.25$~ML coverage) on Au(110)-(1 $\times$ 1) at (a) atop site,
(b) bridge site, and (c) face-centered cubic (fcc) site; and on Au$(110)$-$(2\times 1)$ at (d) atop site,
(e) bridge site, and (f) fcc site.  Each top view and side view are illustrated at
the top and bottom panels, respectively.
Color legend: C, dark brown; Au, gold and light yellow.
For Au(110)-(1 $\times$ 1), total-energy calculations indicate that a single C atom adsorbs
preferentially at the bridge site (b), slightly more energertically favorable than at the fcc
site (c) by 0.14 eV, and at the atop site (a) by 1.686 eV. Four-fold C coordination at the
bridge site exhibits C-Au bond distances of 2.06 and 2.11 ~\AA, facilitating metal to C charge
transfer. Similar to the (1 $\times$ 1) structure, the most stable adsorption site for (2 $\times$ 1) is the
bridge site (e) with four-fold coordinated C showing bond lengths of 1.99 and 2.10 ~\AA. The
atop site (d) is the least favorable by 2.06 eV with a similar C-Au bond of 1.84 ~\AA , while
adsorption at the fcc site (f) is slightly more favorable by 0.377 eV.
}
\end{figure}

Selected atomic adsorption sites for a single C adatom per
supercell (i.e. $\theta=0.25$~ML coverage) on Au$(110)$-$(1\times 1)$ and
-$(2\times 1)$ surfaces calculated with DFT are shown in
Figs. \ref{fig3}(a, b, c)  and \ref{fig3}(d, e, f), respectively. For Au$(110)$-$(1\times 1)$,
total-energy calculations indicate that a single C atom adsorbs
preferentially at the bridge site ($E=-77.337$~eV)
[Fig. \ref{fig3}(b), slightly more energertically favorable than
at the face-centered cubic (fcc) site ($E=-77.194$~eV) [Fig. \ref{fig3}(c)],
and at the atop site ($E=-75.651$~eV) [Fig. \ref{fig3}(a)]. Four-fold C
coordination at the bridge site exhibits C--Au bond distances of
2.06 and 2.11~\AA, in order for C to reach the electronic structure of
Ne:[He]$2s^22p^6$ by metal to C charge transfer. The C adatom at the
fcc site features a distorted four-fold coordination with C--Au bonds
between 2.07 and 2.47~\AA, while the unique C--Au bond at the atop site
is only 1.84~\AA~long.
For Au$(110)$-$(2\times 1)$, the atop site is also the least favorable
($E=-69.424$~eV) with a similar C--Au bond of 1.84~\AA~[Fig. \ref{fig3}(d)],
while adsorption at the fcc site ($E=-71.107$~eV) is slightly more favorable,
with five-fold C coordination and a C--Au bond length of $=2.10-2.20$~\AA, accompanied
by a drastic reconstruction of the top-lying Au layers [Fig. \ref{fig3}(f)].
Similar to the $(1\times 1)$ structure, the most stable adsorption site
for $(2\times 1)$ is the bridge site ($E=-71.484$~eV) with four-fold
coordinated C, showing bond lengths of 1.99 and 2.10~\AA~[Fig. \ref{fig3}(e)].


\subsection{Carbon diffusion on Au$(110)$}

\begin{figure}[b]
\begin{center}
\includegraphics[width=0.5\textwidth]{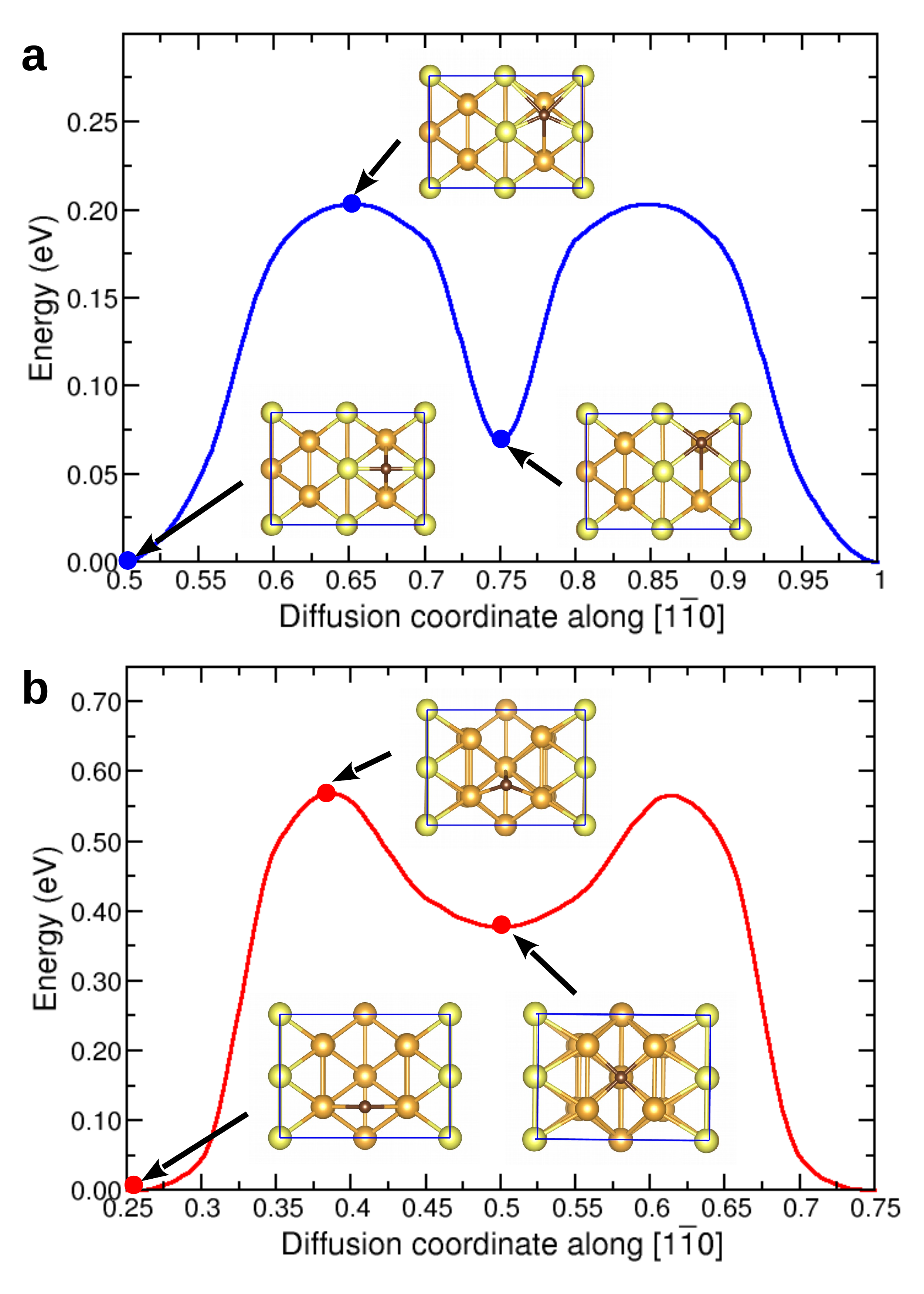}
\end{center}
\caption{\label{fig4} Calculated carbon adatom diffusion path on
(a) Au$(110)$-$(1\times 1)$ and (b) Au$(110)$-$(2\times 1)$ along the $[1\overline{1}0]$ direction
($\theta=0.25$~ML coverage).
Color legend: C, dark brown; Au, gold and light yellow.  Starting from the most energetically favorable bridge sites for C adsorption
on the Au(110)-$(1\times 1)$ and -$(2\times 1)$, the energetics of the C diffusion pathways along the $[1\overline{1}0]$
direction was computed at $\theta$ = 0.25 ML coverage. The predicted energy barrier heights $E_a$
for C diffusion from the bridge site are 0.20 and 0.57 eV for $(1\times 1)$ and $(2\times 1)$, respectively.
The fcc site corresponds to a local energy minimum located 0.07 and 0.37 eV above the
bridge site for $(1\times 1)$ and $(2\times 1)$, respectively (i.e., local C diffusion barriers of $E_a$ = 0.13
and 0.20 eV surround the fcc site for those structures).
}
\end{figure}

Starting from the most energetically favorable bridge sites for
C adsorption on the Au$(110)$-$(1\times 1)$ [Fig. \ref{fig3}(b)]
and -$(2\times 1)$ [Fig. \ref{fig3}(e)] surface structures,
the energetics of the C diffusion pathways along the
$[1\overline{1}0]$ direction was computed in the thermally-activated
regime [Figs. \ref{fig4}(a) and (b)] for $\theta=0.25$~ML coverage.
The predicted energy barrier heights, $E_a$, for C diffusion from the bridge
site are 0.20 and 0.57 eV for $(1\times 1)$ and $(2\times 1)$, respectively.
The fcc site corresponds to a local energy minimum located 0.07 and
0.37 eV above the bridge site for $(1\times 1)$ and $(2\times 1)$,
respectively (i.e., local C diffusion barriers of $E_a=0.13$ and 0.20 eV
surround the fcc site for those structures).


\begin{figure*}
\begin{center}
\includegraphics[width=1.0\textwidth]{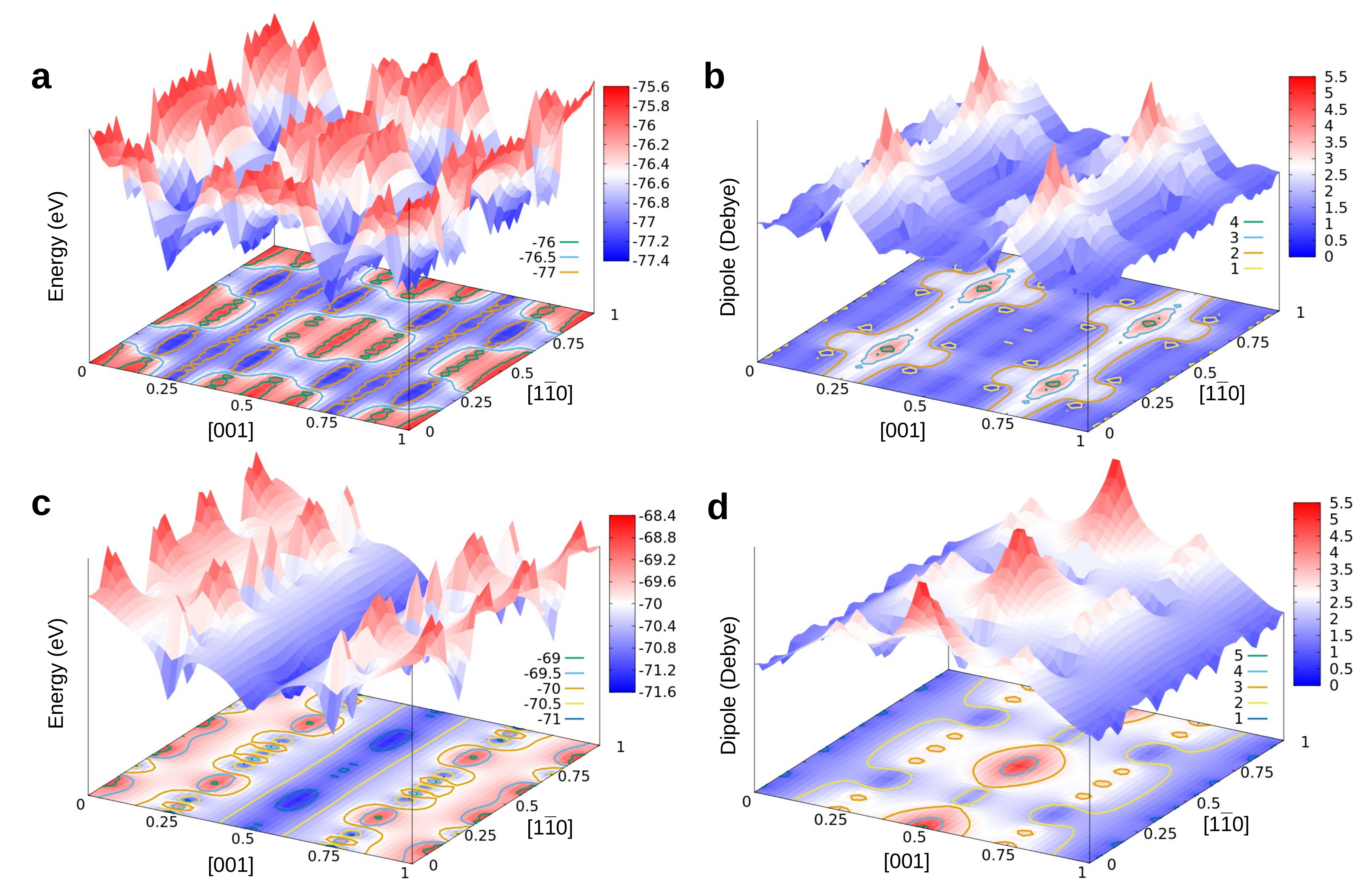}
\end{center}
\caption{\label{fig5} (a) Calculated energy landscape and (b) dipole
map for a carbon adatom diffusing on the Au$(110)$-$(1\times 1)$ surface and (c) energy landscape and (d) dipole map on Au$(110)$-$(2\times 1)$ surface,
along the $[001]$ and $[1\overline{1}0]$ directions at $\theta$  = 0.25 ML carbon coverage. The diffusion path in the
 energetically favorable trench along the $[1\overline{1}0]$ direction, plotted in
 Fig. 4, can be seen. At the bottom of each figure is a contour plot of
 the energy and dipole 3D maps; the contour line values are given on
 the right vertical axis in each figure and are color coded.}
\end{figure*}


Using partitioning of a charge density grid into Bader volumes,
a charge analysis was carried out to assess the charge transfer
occurring at the different C adsorption sites shown in Figs.
\ref{fig4}(a) and (b).
The change in the surface electric dipole $\mu$ along the surface normal
upon C adatom adsorption was obtained using the expression
$\Delta \mu=d_z.\Delta q.e$, where $\Delta q.e$ is the charge transfer
between the surface and the C adatom, and $d_z$ is the $z$-component of the
distance between the C adatom and the surface.

For Au$(110)$-$(1\times 1)$, the predicted $\Delta \mu$ was 2.9, 3.5 and
5.3 D (1 Debye = $3.336 \times 10^{-30}$ Cm) for C adsorption at the bridge, top of the diffusion barrier and
fcc sites, respectively (Fig. \ref{fig4}(a)).
For $(2\times 1)$, $\Delta \mu$ was 2.0, 4.2 and 5.4 D for C adsorption
at the bridge, top of the diffusion barrier and
fcc sites, respectively (Fig. \ref{fig4}(b)).

Calculations were extended to the diffusion of a single carbon adatom
on the Au$(110)$-$(1\times 1)$ and Au$(110)$-$(2\times 1)$
surfaces along the $[001]$ and
$[1\overline{1}0]$ directions ($\theta=0.25$~ML coverage).
The resulting energy landscape and dipole maps for C diffusion on
Au$(110)$-$(1\times 1)$ and Au$(110)$-$(2\times 1)$ are shown in
Fig. \ref{fig5}. These energy maps
confirm that the C adatom diffusion pathway along the $[1\overline{1}0]$
direction displayed in Fig. \ref{fig3} is the most
favorable, while significant diffusion barriers exist along the [001] direction
between adjacent $\left<1\overline{1}0\right>$ rows. As shown in the dipole
maps,
this energetically-favorable diffusion pathway along $[1\overline{1}0]$
also produces the largest values of surface dipole moment, owing to the
efficient charge transfer from the Au surface to the C adatom. For both
$(1\times 1)$ and $(2\times 1)$, the largest surface dipole is obtained
for C occupying the fcc site, corresponding to a local energy minimum between
bridge sites.



\subsection{Diffusion model of motional heating in ion traps}

In this section we present a discussion on the diffusion mechanism which may lead to motional heating of the trapped ion. Surface adsorbates can lead to the motional heating
of the trapped ion through different mechanisms. In particular, if the adsorbates are stationary, the fluctuations in the magnitude of the induced dipole moment $\mu$ is facilitated by the coupling to the phonon modes
of the trap surface~\cite{Safavi-Naini11}. In the opposite limit, mobile adsorbates diffusing on the surface change the magnitude and spatial distribution of the dipoles on the surface, which in turn contributes to the electric field noise.

The role of diffusing particles in the generation of field noise has been studied in the context of field-emission microscopy~\cite{Kleint, Gomer, GesleySwanson}, where adatom diffusion creates fluctuations in the field-emission current in the probed region. More recently this formalism has been applied to explain the motional heating observed in planar ion traps~\cite{Hite12, Rabl}.
Two observations in our experimental system point to the role of diffusion in the motional heating of the ions, first is the scaling of the electric-field fluctuation spectrum with trapping frequency \cite{Hite12}, and the second is the suppression of the noise with decreasing surface adsorbate concentration.

We start by presenting a brief summary of the formalism developed in~\cite{Rabl} and provide estimates for the diffusion noise spectral density using the DFT parameters.
A diffusion adatom is represented by a surface polarization density $P(\vec r, t) = \mu \sigma(\vec r, t)$, where $\sigma$ is the areal density of adatoms, and $\mu$ is the dipole moment of the adsorbate.  This creates an electric-field noise spectrum
$$
S_E= \frac{\mu^2}{8\pi^2 \epsilon_0^2} \int_S d^2 r_1 \int_S d^2 r_2 g_D(\mathbf{r}_1)g_D(\mathbf{r}_2) C_\sigma(\mathbf{r}_1,\mathbf{r}_2, \omega)
$$
where $g_D(\mathbf{r})=\frac{2d^2 -r^2 }{\vert d^2+r^2\vert^{5/2}} \left(\frac{3dx }{\vert d^2+r^2\vert^{5/2}}\right)$ , a geometric factor for the dipole pattern, is for electric-field fluctuations perpendicular (parallel) to the trap surface, with $\mathbf{r}=(x,y)$, the position of the adatom on the trap surface and $d$, the distance between the surface of the electrode and the trapped ion.
Here
$$
C_\sigma(\mathbf{r}_1, \mathbf{r}_2, \omega)=2{\rm Re} \int_0^\infty d\tau \langle \delta \sigma(\mathbf{r}_1,\tau) \delta \sigma(\mathbf{r}_1,0) \rangle e^{-i \omega t}
$$
is the correlation spectrum of the density fluctuations, $\delta\sigma(\vec r, t) = \sigma(\vec r, t)-\langle \sigma (\vec r, t)\rangle$.
For low densities of adsorbates, the adatoms can be modeled as independent particles diffusing over the trap surface with
$$
\langle \delta \sigma(\mathbf{r}_1,\tau) \delta \sigma(\mathbf{r}_1,0) \rangle= \frac{\bar \sigma}{4 \pi D \tau} e^{- \frac{\vert \mathbf{r}_1- \mathbf{r}_2\vert^2}{4 D \tau}}
$$
as shown in~\cite{GesleySwanson}. In the above expression, ${\bar \sigma}$ is the stationary value of $\sigma(\vec r, t)$ in the case of a homogeneous surface. This expression can be used to obtain analytic expression for the electric-field fluctuation spectrum specific to a variety of trap geometries as shown in ~\cite{Rabl}. However, the resulting electric-field fluctuation spectrum for all geometries, if attributed to adatom diffusion, is many orders of magnitude too small to explain the observed noise in the ion traps.

One can further refine this model by incorporating the fluctuations of the magnitude of the induced dipole moment as the dipoles move between patches of varying work functions, such as those described in the previous sections. In this case, the noise resulting from diffusion over each patch with lateral dimensions $R_{\rm el}$ is modeled by the noise due to a small electrode ($R_{\rm el}\ll d$) ~\cite{Rabl},
\begin{equation}
\label{eq:SEnopatch}
S_{E, \perp}\approx \frac{\mu^2 \bar \sigma R_{\rm el} \sqrt{D}}{\sqrt 2 \pi \epsilon_0^2 d^6 \omega^{3/2}}.
\end{equation}
The most important distinction between the patch model and diffusion over the electrode surface is the change in the induced dipole moment of the different patches, as well as the enhancement due to the number of patches in the probed area $N_{\rm p}\sim d^2 /R_{\rm p}^2$. Using the results of Ref.\cite{Rabl}, we find that the electric dipole fluctuation spectrum, taking into account the surface patches, is given by

\begin{equation}
\label{eq:SEpatch}
S_{E, \perp}\approx \frac{\Delta \mu^2 \bar \sigma  \sqrt{D}}{\sqrt 2 \pi \epsilon_0^2 d^4 \omega^{3/2}R_{\rm patch}}
\end{equation}
where $\Delta \mu$ is the fluctuation in the induced dipole moment between the patch and the clean trap surface.

In order to make the above expression more realistic, we point out that we have assumed that the diffusion constant over the trap surface does not change dramatically. However, if there is a large energy barrier between the patch and the trap surface, then the motion of the particle onto the patch will be hindered. Moreover, the transfer rate to the patch is affected by the surface coverage. One can account for this effect by constructing a simple model, described in ~\cite{Kleint}.

Let us assume that the adsorbates, at steady state, can move between sites of type ``S" or ``P" separated by an energy barrier $\Delta E$, with ``P" sites sitting at lower energies. The surface consists of $N$ sites, with $N_S$ and $N_P$ type ``S" and ``P" sites, respectively. Here the transition between the two types of sites are driven by energetic excitations, and the adsorbates are non-interacting. The mean values of the adsorbate concentrations on each site, are given by the rate equations
\begin{align*}
\frac{d(N_S/N)}{dt}&=\dot n_S= \alpha_{PS} n_P- \beta_{SP} n_S(N_P-n_P)\\
\frac{d(N_P/N)}{dt}&=\dot n_P= -\dot n_S
\end{align*}
where $\alpha_{PS}$ and $\beta_{SP}$ are the transition coefficients between sites P and S. These different sites with differing activation energies correspond to different adsorption centers, each with adsorbate
 surface concentration $n_P$ and $n_S$, respectively. The total number
 of sites is $N=N_P+N_S$.

 In this scenario, the number of transitions per unit area and time between the surface and the patch is given by
\begin{equation}
K=\beta_{SP} N_P N_S \theta (1-\theta) e^{-\Delta E/k_b T_{eff}}
\end{equation}
where $T_{eff}$ is an effective temperature to account for any driving mechanisms, and $0 \leq \theta \leq 1$ is the surface coverage. This simple formula illustrates how the number of transitions between the two types of sites, which is crucial to the patch model of the diffusion noise described above, goes to zero at both $\theta=0$ and 1, and is maximized at some intermediate filling depending on the ratio of sites belonging to each type, and temperature. {The non-monotonic behavior of $K(\theta)$ with decreasing surface coverage is qualitatively} consistent with the variation in the size of the electric field noise observed in the trap used in our setup, as well as other electric field noise observations related to surface diffusion mechanisms. {Furthermore, one can use $K$ to refine the expression for the noise spectral power in Eq.~\eqref{eq:SEpatch} to account for the different rates of surface diffusion of the particles in a single patch ($D$) or in between patches ($K$). }

The above phenomenological model illustrates the variation of electric-field noise with physical parameters. It also helps to explain the wide variations in measured values across the literature.  In this work, we make a direct comparison of experimental and theoretical contributions to the noise.  This provides an order-of-magnitude estimate for the size of the electric-field noise attributable to surface diffusion over an imperfect surface.

Using the experimentally determined parameters $R_{\rm patch}\sim 0.1-1\mu$m and $d \sim 40-100\mu$m in combination with the results of the calculations presented above for $\Delta \mu\approx 2-5$~D, we estimate the size of the electric-field noise attributable to the diffusion of independent adsorbates. The typical range of the diffusion constant at room temperature can be estimated as $D\sim10^{-14}-10^{-11}$ m$^2$s$^{-1}$ for barrier heights between $200-500$ meV. Since the assumption of independent adsorbates is valid for low surface coverages, we use $\bar \sigma \sim 10^{18} m^{-2}$, and  find  $S_{E, \perp}(\omega_t=1~{\rm MHz})\approx 10^{-15}-10^{-10}$~ V$^2$/m$^2$ Hz, comparable to measured experimental values.

\begin{figure}[t]
\begin{center}
\includegraphics[width=0.45\textwidth]{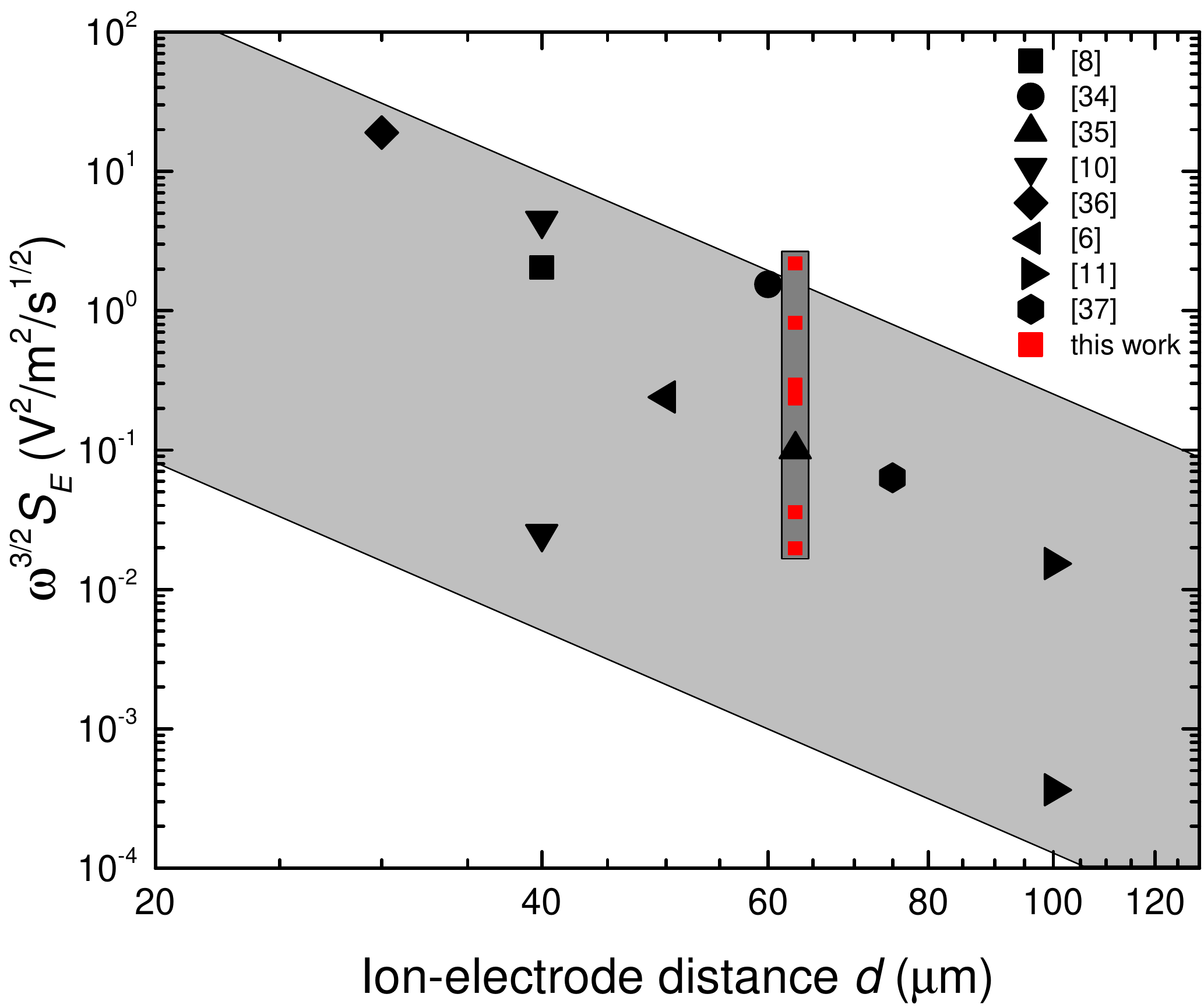}
\end{center}
\caption{\label{fig6} Measured frequency normalized electric-field noise spectral densities for the lowest noise room-temperature surface-electrode ion traps plotted as a function of the ion-electrode distance.  These data are overlaid on top of our estimated range from Eq. (4), shown as a light grey band, and reasonable values for the diffusion constant and patch size.  Data from Fig. 1 are also plotted here in red for comparison.
}
\end{figure}

In Fig. \ref{fig6}, we construct a plot of the frequency normalized electric-field noise spectral density $\omega^{3/2}S_E$ vs. $d$ the distance from the surface of an ion-trap electrode.  We have compiled a set of data from the literature of various room-temperature surface-electrode traps, and overlaid them on top of our estimated range (light grey band) for this type of noise.  We also include for comparison the data from Fig. \ref{fig1-noisedata} (red points with dark grey band).  The estimated range from our calculations agrees well with data from the literature.  In fact, when considering the parameters in Eq. 4, and the known quantities from the electric-field noise data shown in this work (cf. Fig. \ref{fig1-noisedata}), the only two poorly known values are the diffusion constant $D$ and the patch size $R_{patch}$.  These can be combined in terms of a
 diffusion time constant, $\tau = R_{patch}^2/D$.  Evaluation of Eq. 4 to find values of $\tau$ for the data in Fig. 1 yields $\tau \approx$ 8.5 ms for the highest noise spectral density and more than 1 s for the low noise spectral density.

\section{Summary and Outlook}

In summary, we have shown new experimental data of  motional heating
in a stylus-type ion trap, and used first-principles DFT calculations
to compute the energy landscapes and dipole maps for a carbon adatom
diffusing on the Au(110)-(1$\times$1) and -(2$\times$1) surfaces. In light of these
results, we discuss how the fluctuating dipole moment from a diffusing
carbon adatom is a possible source of motional heating in ion traps, and
compute an estimated range for the electric-field noise spectral density.
These data agree well and give further insight into the origin of
anomalous heating in ion traps.

A crucial parameter in the precise determination of the diffusion noise is
the diffusion rate and how it varies with temperature. In the  future,
using long-time molecular dynamics simulation of carbon diffusion
on Au(110) surfaces, we aim to obtain values for the diffusion constant.
We will also aim to better constrain the spread in the calculation of the
diffusion noise. The numerical simulations will be extended to energy and
dipole moments of adsorbed hydrocarbon molecules. The
resulting noise spectrum due to  such molecular
species on Au surfaces will be calculated. On the experimental side,
we are presently conducting experiments to measure the diffusion time
constants for these surfaces independently using tunnel-current
fluctuation measurements with STM.

\subsection*{Acknowledgments}

The authors thank D. Wineland, D. Leibfried and
S. Kotler for helpful suggestions on the manuscript.  AS and HRS benefitted from discussions with P. Rabl.  Sandia National Laboratories is a multiprogram laboratory operated
by Sandia Corporation, a wholly owned subsidiary of Lockheed Martin
Company, for the United States Department of Energy's National Nuclear
Security Administration under Contract DE-AC04-94AL85000.  This article is a
contribution of the U.S. Government and is not subject to
U.S. copyright.




\begin{thebibliography}{}
\bibitem{Footnote1} Occasionally, small levels of sulfur and/or oxygen are detected with AES on the as-fabricated trap chips.  Interestingly, these contaminants (S and O) are typically removed after the first low dose sputter treatment, while the carbon remains present until higher doses.
\end{thebibliography}
\end{document}